\documentclass[summary]{ursi}
\usepackage[utf8]{inputenc}

\title{Exploring the Use of Generative AI in the Search for Extraterrestrial Intelligence (SETI)}

\author{John Hoang *\affref{ref1}, Zihe Zheng\affref{ref2}, Aiden Zelakiewicz\affref{ref3}, Peter Xiangyuan Ma\affref{ref4}, and Bryan Brzycki\affref{ref5} }

\affiliation{%
  \aff{ref1}{SCIPP, California, USA; e-mail: jokhoang@ucsc.edu}
  \aff{ref2}{Yale University, Connecticut, USA; e-mail:  zihe.zheng@yale.edu}
  \aff{ref3}{The Ohio State University, Ohio, USA; e-mail:  zelakiewicz.1@osu.edu}  
  \aff{ref4}{Department of Mathematics, University of Toronto; e-mail: peterxy.ma@mail.utoronto.ca}
  \aff{ref5}{Department of Astronomy, University of California Berkeley; e-mail: bbrzycki@berkeley.edu}
}

\begin{document}

\maketitle

\begin{abstract}
The search for extraterrestrial intelligence (SETI) is a field that has long been within the domain of traditional signal processing techniques. However, with the advent of powerful generative AI models, such as GPT-3, we are now able to explore new ways of analyzing SETI data and potentially uncover previously hidden signals. In this work, we present a novel approach for using generative AI to analyze SETI data, with focus on data processing and machine learning techniques. Our proposed method uses a combination of deep learning and generative models to analyze radio telescope data, with the goal of identifying potential signals from extraterrestrial civilizations. We also discuss the challenges and limitations of using generative AI in SETI, as well as potential future directions for this research. Our findings suggest that generative AI has the potential to significantly improve the efficiency and effectiveness of the search for extraterrestrial intelligence, and we encourage further exploration of this approach in the SETI community. (\textbf{Disclosure}: For the purpose of demonstration, the abstract and title were generated by ChatGPT and slightly modified by the lead author.)
\end{abstract}

\section{Introduction}
The Breakthrough Listen project has been searching for technosignatures in our universe using powerful radio telescopes around the world, including the Green Bank Telescope, Parkes Telescope, Allen Telescope Array, and MeerKAT. The most popular technique in radio SETI involves looking for narrow-band signals in the time-frequency spectrogram data (often referred to as dynamic spectra or waterfall plots in existing literature). In recent years, machine learning (ML) algorithms have been employed to classify image-like spectrograms~\cite{PeterMa}, and \texttt{setigen}, an open-source Python library, was created to synthesize mock labelled radio SETI training data set ~\cite{Brzycki}. Since \texttt{setigen} is meant to be a general-purpose heuristic framework, there are rooms for improvement if one wishes to improve the speed of the algorithm in specialized cases. The method used in this work is Generative Adversarial Network (GAN)~\cite{Goodfellow}. 

A GAN consists of two competing deep neural networks: the generative network and the discriminator network. The generator generates images from random noise, and the generated images are fed into the discriminator along with real images. The discriminator then classifies the images as either real or fake and both the generator and the discriminator models are updated according to the result. At equilibrium, the generator will generate images that look similar to the training data and as a consequence, the discriminator will no longer be able to distinguish between generated and real images. The output of the generator will determine whether an image is similar to the training data or not, which can be useful in determining whether the image is normal or anomalous. Therefore, GAN can also be used to detect outliers. In astronomy, GANs have been used to simulate galaxy images~\cite{Dia}, gamma-ray Cherenkov airshower signals~\cite{Dubenskaya}, detecting outliers~\cite{Bentabol}, to name but a few. GANs and its family of other generative frameworks such as AutoEncoder are parts of what is colloquially known as DeepFake, a portmanteau of Deep learning and Fake.  

\section{Experiments}
\begin{figure}[ht]
  \centering
  \includegraphics[width=55mm]{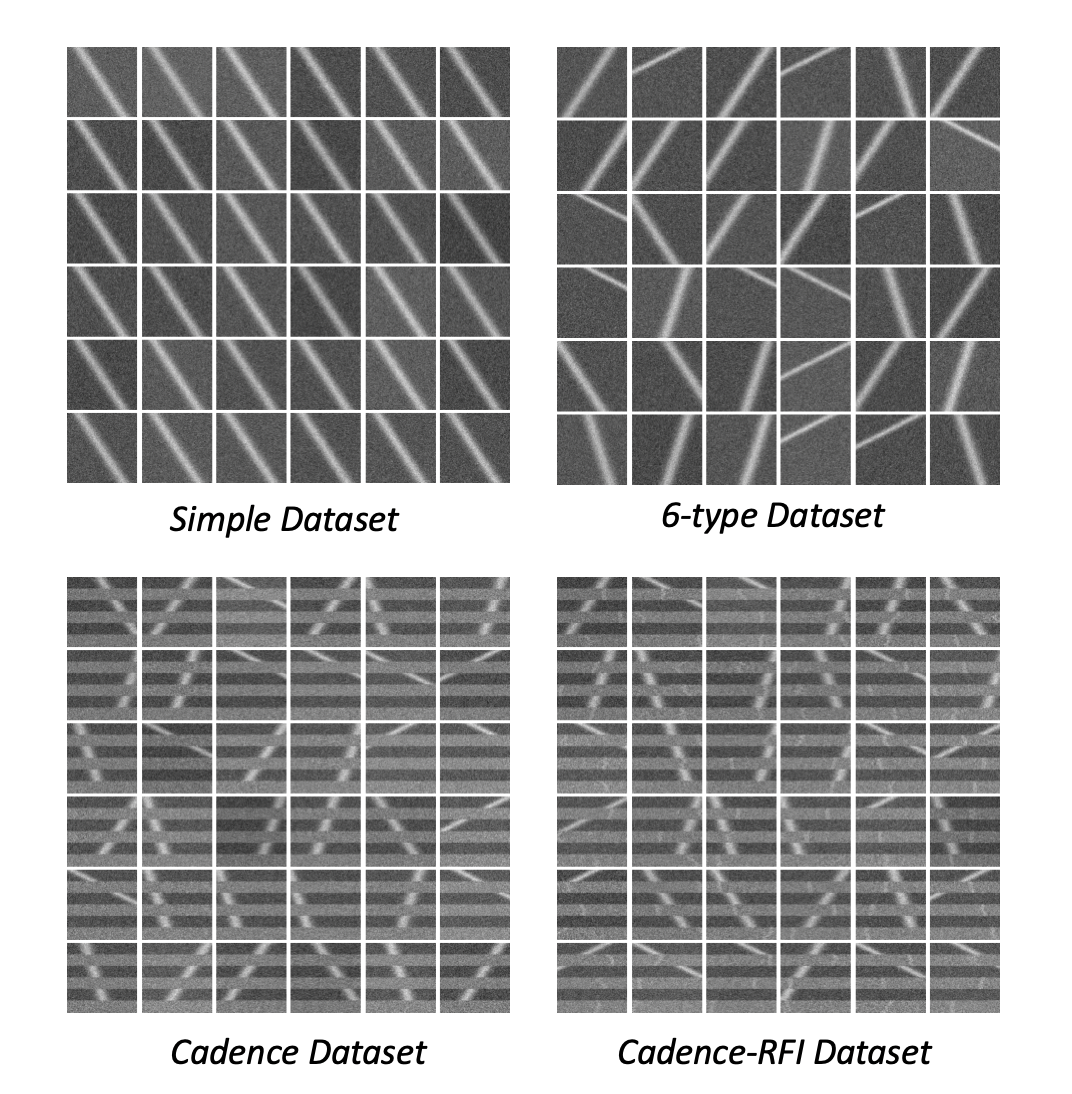}
  \caption{4 narrow-band data sets used for training in this work with increasing complexity. Top row: narrow-band signals with a constant drift rate, narrow-band signals with 6 drift rates. Bottom row: ON-OFF cadence observations with narrow-band ETI-like signals in only ON observation, ON-OFF cadence observations with narrow-band ETI-liked signal in only ON observation and random narrow-band RFI-like in OFF observation.}
  \label{fig:dataset}
\end{figure}

We employ the \texttt{setigen} software to generate waterfall plots used for training. Each generated set includes 15000 waterfall plots, each having 128 pixels in time and 128 pixels in frequency. Background noise is randomly chosen from either Gaussian noise or Chi-squared noise, with mean equals to 10 and standard deviation (only for Gaussian noise) equals to 1. Injected narrow-band signals are specified by the starting point, drift rate and width in each data set. The width of the signals all range from 50 to 60 pixels. Random walk signals are injected in some of the data sets to simulate radio-frequency interference (RFI), which is frequently seen in real observations. These 4 sets used in the experiments are shown in Fig.\ref{fig:dataset}.

\subsection{Simple drifting signals}
\begin{figure}[htbp]
  \centering
  \includegraphics[width=60mm]{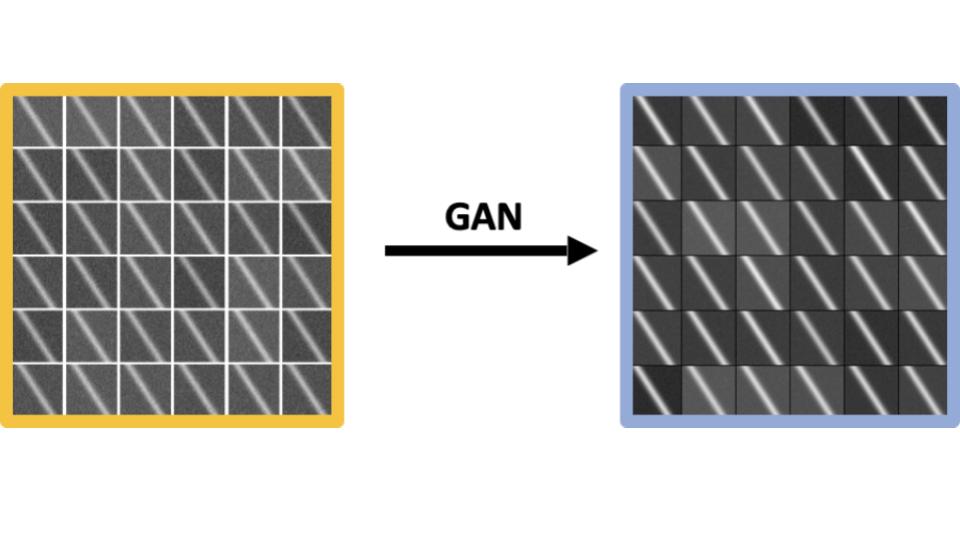}
  \caption{Vanilla GAN is able to reproduce the training data from \texttt{setigen} (left): right plot contains GAN-generated waterfall plots with random background noise level}
  \label{fig:Result_1}
\end{figure}

We first test GAN's ability to generate waterfall plots containing narrow-band signals with a single drift rate. As seen from Fig.\ref{fig:Result_1}, GAN successfully reproduces this data set after training.

\subsection{Simple drifting signals with varying drift rates}
We then expand the variety of data by changing the drift rate and starting points of signals in the waterfall plots, which result in the 6-type dataset. We train the same vanilla GAN network architecture on the 6-type data set shown in Fig.\ref{fig:dataset} for 1000 epochs. However, the network fails to produce realistic waterfall plots.

\begin{figure}[htbp]
  \centering
  \includegraphics[width=60mm]{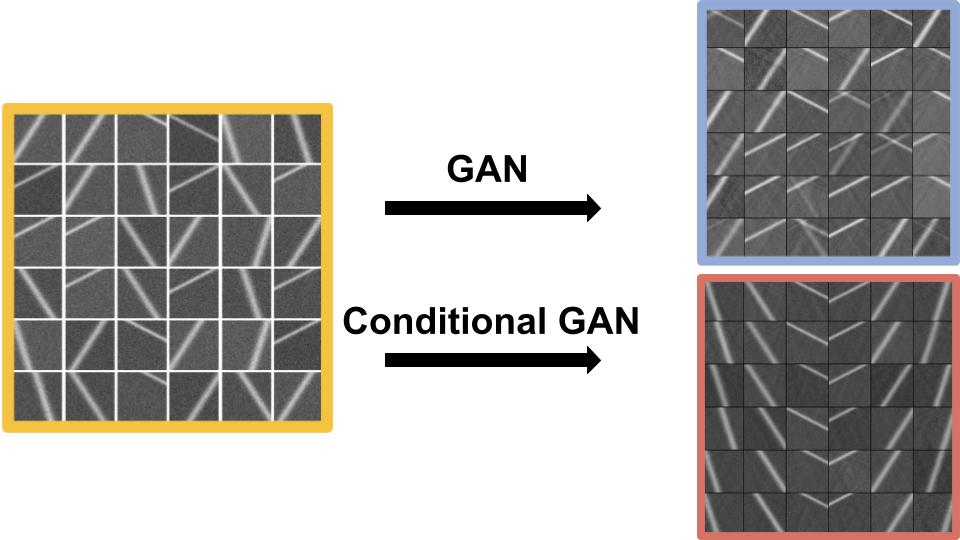}
  \caption{Vanilla GAN struggles to learn the distinct features among different drift rates. However, conditional GAN solves the issue by artificially condition training data on the slope label.}
  \label{fig:Result_2}
\end{figure}

Conditional GAN~\cite{Mirza} is then selected as our next candidate architecture among several variations of GAN since it allows the network to generate images conditioned on the class of data. Conditional GAN achieved this idea by adding labels for data according to their drift rates, which are fed in the network along with the plot themselves. We labeled the 6-type data set with 0, 1, 2, 3, 4, 5 according to 6 different drift rates, and trained the conditional GAN on the data set for 1000 epochs. The generator from the conditional GAN is able to generate waterfall plots with a singular signal with various starting points in each class. The results in comparison with traditional GAN are shown in Fig.\ref{fig:Result_2}.

\subsection{Cadence signals}
\begin{figure}[htbp]
  \centering
  \includegraphics[width=60mm]{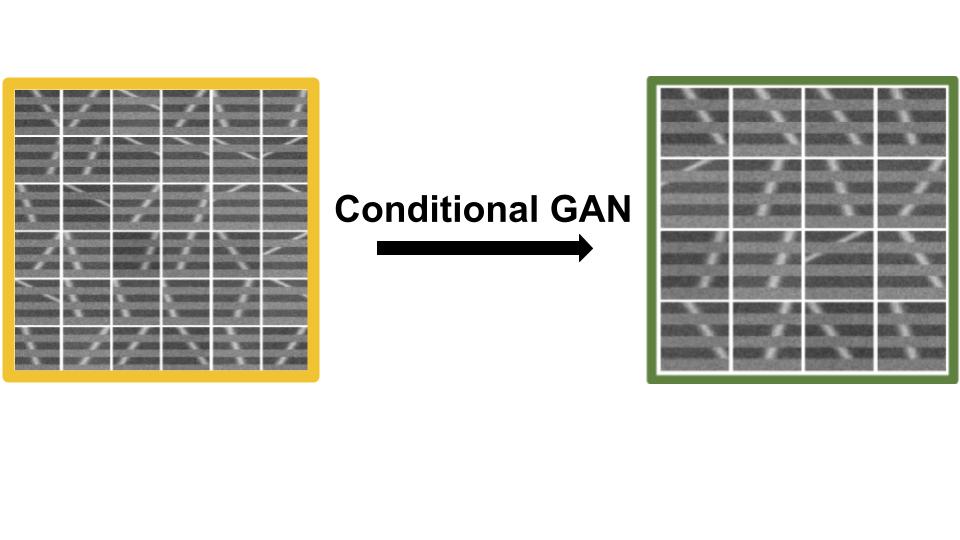}
  \caption{Conditional GAN is able to learn the more complex ON-OFF signal patterns mimicking typical SETI observations.}
  \label{fig:Result_3}
\end{figure}

Due to the prevalence of radio frequency interference, a typical radio SETI observation consists of pointing the telescope alternatively to ON-OFF sky locations, and a potential narrow-band ETI signal candidate should only show up in the ON observations. To simulate this, three horizontal noisy band of size 21*128, 21*128 and 23*128 pixels are added to each waterfall plots to simulate the ON-OFF cadence in real observations. Successful results are shown in Fig.\ref{fig:Result_3}

\subsection{Cadence signals with RFI}
\begin{figure}[ht]
  \centering
  \includegraphics[width=60mm]{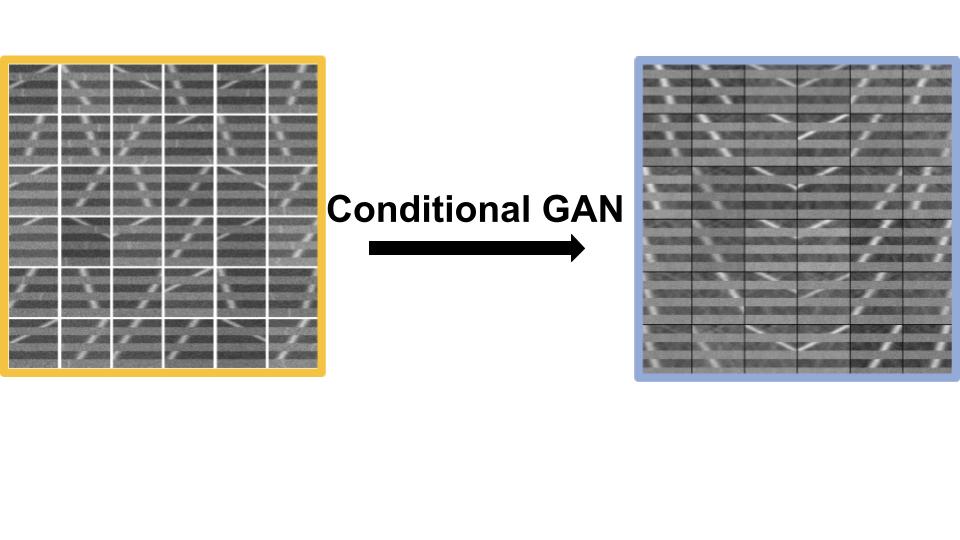}
  \caption{Conditional GAN is able to ignore RFI in the training data. For clarity, in the right plot, waterfall plots in each column have the same label (drift rate).}
  \label{fig:Result_4}
\end{figure}

We finally test our generator on its resistance to RFI in the dynamic spectra. We trained our conditional GAN model on the Cadence-RFI dataset, which contains injected RFI noise. The result is shown in Fig.\ref{fig:Result_4}. The generator ignores the presence of RFI and does not produce them in the OFF observations. In addition, the generator also avoids a common failure mode called mode-collapse whereby GAN only learns to output a particularly plausible output and nothing else. Fig.\ref{fig:Result_4} shows a degree of variability within a single class: although all the waterfall plots in the same column have the same drift rate (label), they vary with intensity and location within the frame.

\subsection{Generator's scores}
\begin{figure}[ht]
  \centering
  \includegraphics[width=75mm]{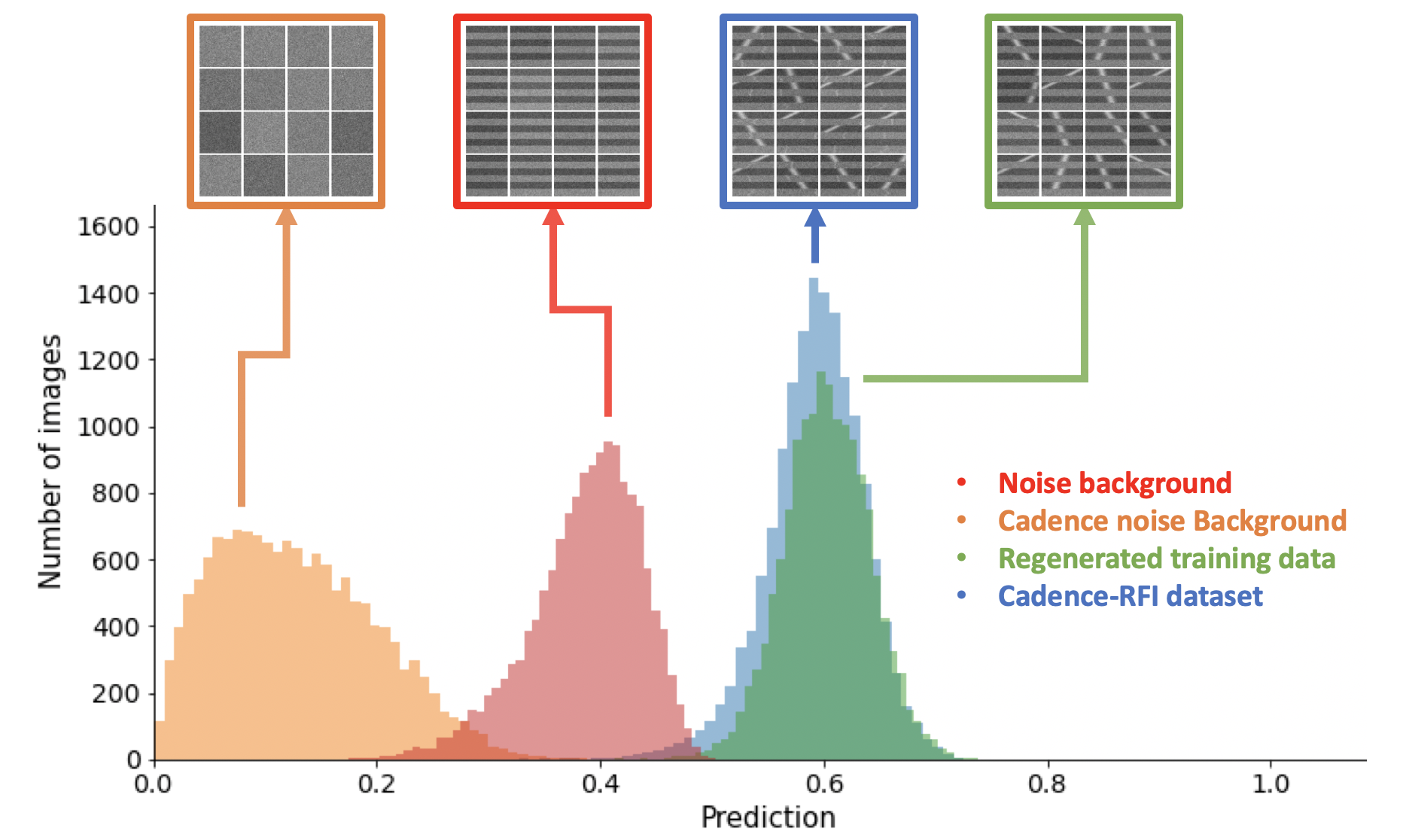}
  \caption{Distribution of scores for different data sets by the discriminator.}
  \label{fig:Result_5}
\end{figure}
Because the discriminator learns to classify realistic waterfall plots and unrealistic waterfall plots during training, the trained discriminator can be used subsequently as a classifier. We train our conditional GAN on the cadence data set in Fig.\ref{fig:dataset}. To test the trained discriminator, we used the 4 data sets shown in Fig.\ref{fig:Result_5}, each containing 15000 waterfall plots: only noise background, only cadence background, GAN-produced data, and the cadence-RFI data. It is clear that waterfall plots that are less similar to training data receive lower scores from the discriminator than those are similar to the training data set. The blue and green distribution have similar means, confirming that both generator and discriminator disregard RFI in the training data.

\subsection{Timing}
To illustrate the advantage of using pre-train model, we show in Tab.\ref{tab:Timing_table} the time it took our GAN to generate and write to storage $10^2$, $10^3$, $10^4$, $10^5$ waterfall plots. Comparing with the clearly linear time of \texttt{setigen}, GAN's speed is superior when a large data set is involved, and we are only limited by the availability of GPU RAM and storage I/O.

\begin{table}[]
\caption{CPU/GPU time to generate (G) and write (W) to storage as a function of number of waterfall plots, using both heuristic \texttt{setigen} with CPU and generative GAN with GPU. GAN excels when a large number of waterfall plots are involved, improving the speed by more than 2 orders of magnitude.}
\label{tab:Timing_table}
\begin{tabular}{@{}llllll@{}}
                      &          & \multicolumn{2}{c}{\textbf{SETIGEN}} & \multicolumn{2}{c}{\textbf{GAN}} \\
\multicolumn{1}{c}{\#} &
  \multicolumn{1}{c}{File Size} &
  \multicolumn{1}{c}{G (s)} &
  \multicolumn{1}{c}{G\&W (s)} &
  \multicolumn{1}{c}{G (s)} &
  \multicolumn{1}{c}{G\&W (s)} \\
$10^2$ & 12.5 MB  & 0.578             & 0.613            & 3.335          & 3.378           \\
$10^3$ & 125.3 MB & 5.09              & 6.287            & 3.784          & 4.181           \\
$10^4$ & 1.2 GB   & 51.279            & 55.723           & 3.807          & 7.724           \\
$10^5$ & 12.2 GB  & 515.802           & 563.833          & 4.358          & 95.597         
\end{tabular}
\end{table}

\section{Possible extensions}
\begin{figure}[ht]
  \centering
  \includegraphics[width=60mm]{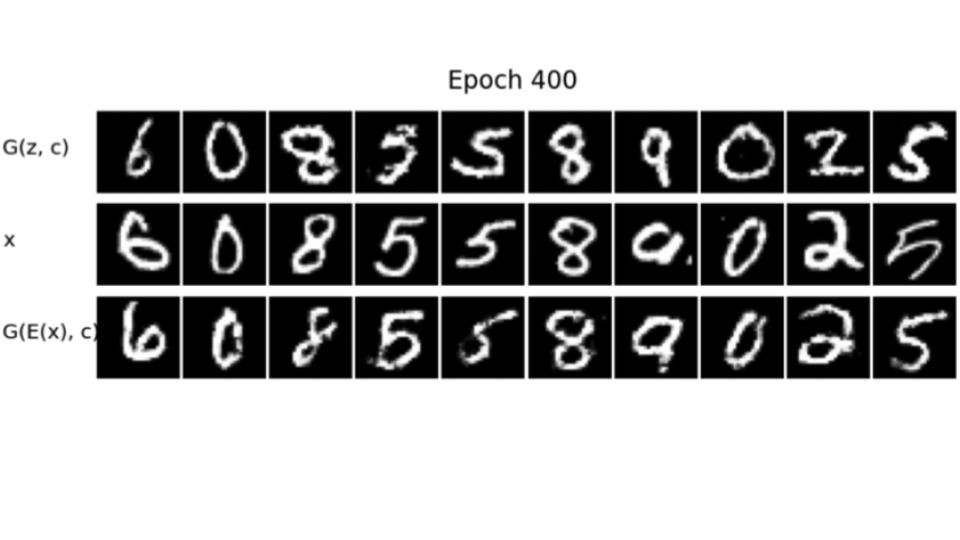}
  \caption{MNIST images synthesized by a BiCoGAN's generator with random latent vectors (top row), comparing with original training data (middle row) and with reconstructed latent vectors from the encoder network (bottom row).}
  \label{fig:BiCoGAN}
\end{figure}

 It is possible to further modify GANs into a more advanced bi-directional architecture so that we can reverse-engineer images into meaningful latent vectors. The proposed Bidirectional Conditional GAN (BiCoGAN)~\cite{Jaiswal} includes an additional encoder network that is trained simultaneously with the generator and the discriminator, and can provide inverse mappings of data samples to both intrinsic and extrinsic vectors. The benefit of BiCoGAN is two-fold: the reverse-engineered latent vectors of hitherto unseen signals can be used to compare with those of known signals to identify anomalies, and anomalous latent vectors can be slightly perturbed and subsequently fed into the generator to synthesize new anomalous signals useful for further training. This remains a work in progress, and an example of hand-written digits generated by BiCoGAN is shown in Fig.\ref{fig:BiCoGAN}.

\section{Limitations}
\begin{figure}[ht!]
  \centering
  \includegraphics[width=90mm]{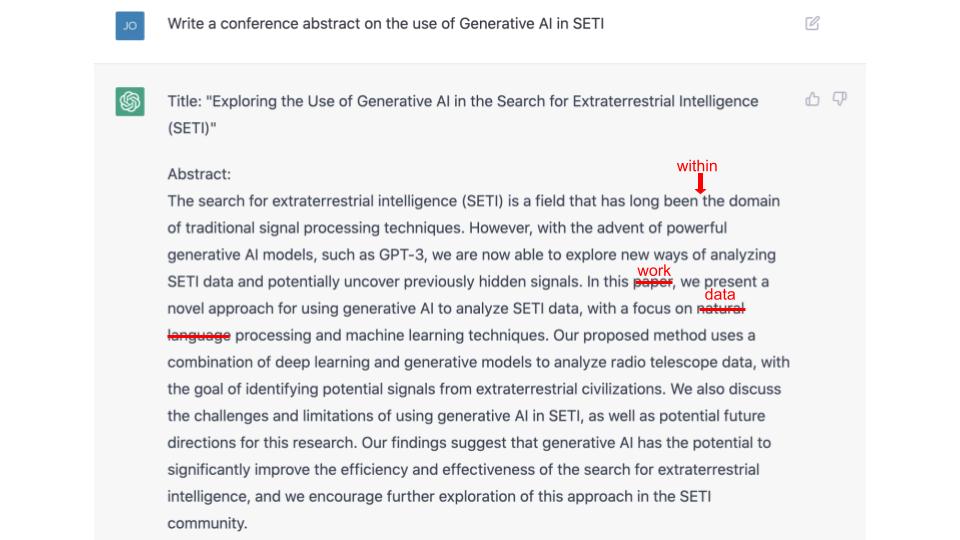}
  \caption{Response by ChatGPT from a prompt by the author, and with  subsequent edits to either improve language usage or correct factual knowledge.}
  \label{fig:ChatGPT}
\end{figure}

For the purpose of demonstrating the limitations of Generative AI, let us examine this paper's title and abstract, which were largely produced by ChatGPT - Chat Generative Pre-trained Transformer. While reads convincingly at first, the generated abstract certainly contains errors. Notably, the relationship between natural language processing and radio data analysis makes little sense, if at all. Nonetheless, upon a handful of minor edits by an expert, the hybrid version of the abstract sounds even more convincing (see Fig.\ref{fig:ChatGPT}). In a similar vein, radio spectra produced by a GAN are realistic overall, but they require more careful benchmark when high precision is required or when statistics are low.

Fundamentally, by the virtue of the Universal Approximation Theorems, neural networks are functional approximator (often polynomial approximation if simple activation functions are used) of a multidimensional distribution to arbitrary precision. In essence, it is analogous to the Weierstrass Approximation Theorem, which is as follows: Suppose f(x) is a continuous real-valued function defined on the real interval [a,b], then: 
\begin{equation}
  \label{eq:WAT}
\forall \epsilon > 0, \exists \ p(x) \quad \mathrm{s.t.} \quad
\forall x \in \mathclose[a,b\mathclose] \,\ , \,\ |f(x) < p(x)| < \epsilon
\end{equation}
In the case of ChatGPT or any Generative AI, the interval [a,b] is the domain in which the training data f(x) is contained. Since the domain of the function is a specific \textit{closed interval} instead of the entire real number line, the generator can approximate or interpolate very well the distribution \textit{within the domain} using a polynomial p(x). In other words, the output is a sufficiently good polynomial approximation of the training data contained therein. In the case of this work's abstract, since ChatGPT most likely obtained training data from publications, it outputs "paper" instead of the more naturally sounded "work" in the context of an abstract for a conference presentation. However, \textit{outside the domain}, the \textit{extrapolated} output is sometimes only tangential, and requires further scrutiny. The ideal training data set should includes both theory-based data as well as real-life examples, although this is hardly achievable in many cases whereby new phenomenon are being investigated. 

\section{Conclusions}
We have explored the potential of deep generative networks for generating dynamic spectra and their ability to detect outliers. The main advantage of such generative AI is that they can encode the complex representations of the training data and rapidly decode them at a rapid speed by leveraging GPU usage. The work is relevant not only to  radio SETI, and can be readily extend to SETI search in other parts of the electromagnetic spectrum. Furthermore, since the AI models used in this work are still far from the industrial grade models such as one used by ChatGPT, there is much for further experimentation with more sophisticated AI architecture. However, as in the case with other applications of Generative AI, we caution against the over-reliance on their outputs, as sometimes they are only tangential to the reality at best, and nonsensical at worst. The code used in this work is publicly available at https://github.com/zzheng18/SETIGAN.

\section*{Acknowledgements}
We thank the Breakthrough Prize Foundation and the University of California, Berkeley, for their support. Special thanks to Andrew Siemion, Steve Croft, and Yuhong Chen for several fruitful discussions. This work is funded by the Breakthrough Listen Initiative and the National Science Foundation.

\end{document}